\documentclass{article}
\usepackage{ijcai21}

\usepackage{times}

\usepackage{soul}
\usepackage{url}
\usepackage[hidelinks]{hyperref}
\usepackage[utf8]{inputenc}
\usepackage[small]{caption}
\usepackage{subfigure}
\usepackage{graphicx}
\usepackage{amsmath}
\usepackage{booktabs}
\usepackage{verbatim}
\usepackage{multirow}
\title{Are All Edges Necessary? \\ A Unified Framework for Graph Purification}
\author{
Zishan Gu$^*$
\and
Jintang Li$^*$\And
Liang Chen$^*$
\affiliations
$^*$Sun Yat-sen University
\emails
\{guzsh,lijt55\}@mail2.sysu.edu.cn, chenliang6@mail.sysu.edu.cn
}

\begin{document}

\maketitle

\begin{abstract}
Graph Neural Networks (GNNs) as deep learning models working on graph-structure data have achieved advanced performance in many works. However, it has been proved repeatedly that, not all edges in a graph are necessary for the training of machine learning models. In other words, some of the connections between nodes may bring redundant or even misleading information to downstream tasks. In this paper, we try to provide a method to drop edges in order to purify the graph data from a new perspective. Specifically, it is a framework to purify graphs with the least loss of information, under which the core problems are how to better evaluate the edges and how to delete the relatively redundant edges with the least loss of information. To address the above two problems, we propose several measurements for the evaluation and different judges and filters for the edge deletion. We also introduce a residual-iteration strategy and a surrogate model for measurements requiring unknown information. The experimental results show that our proposed measurements for KL divergence with constraints to maintain the connectivity of the graph and delete edges in an iterative way can find out the most edges while keeping the performance of GNNs. What’s more, further experiments show that this method also achieves the best defense performance against adversarial attacks. 
% Our framework offers a new perspective towards graph structure data and more robust GNNs.

\end{abstract}

\section{Introduction}

% Graph Neural Networks (GNNs) are a type of reliable models working on graph-structure data, which is quite ubiquitous in the real world by representing entities and their relationships through nodes and edges in applications like social networks, knowledge graphs, etc. Among these models, Graph Convolutional Networks (GCNs), which utilize message passing through a neighborhood aggregation function to extract high-level features from a node and its neighborhoods, have achieved the state-of-the-arts performance for a variety of tasks. Recently, while most GNNs (including GCN) tend to consider all of the edges in a graph to train its model, there are more and more attempts trying to left out edges during the training process, which may lead to a deeper and more robust models against adversarial attacks. The successful training and the enhanced performance of these models happen to prove that not all edges are necessary for a task on graphs, especially graphs under well-designed attacks. In this spirit, we further raise the following questions: Can we find out these redundant or deliberately injected edges before the training process? Can GNNs keep their performance without all those deleted edges on clean graphs? Can we enhance the robustness of GNNs against adversarial attacks by deleting these models?

Graph structure data is a type of ubiquitous non-Euclidean data that is pervasive across many real-life applications, such as social networks, recommendation systems, etc \cite{DBLP:conf/ijcai/ChenL0GZ19,chen2018heterogeneous}. % Representing entities and their relationships through nodes and edges, graphs are indeed a highly abstract yet efficient way to summarize information. 
However, we may all have the experience of friending unfamiliar people on Facebook$\footnote{www.facebook.com}$.
%or buying staff we don’t really care for friends and relatives. 
\cite{10.1145/2076732.2076746} uses 102 bots to send friend requests to random users with public accounts, the results of which indicates that almost 20\% of users are willing to friend total strangers, and approximately 60\% of users would friend strangers with only one common friend, which may all lead to links that we probably shouldn't consider when we study the resulted graphs. In the meantime, studies have shown that GNNs are highly sensitive against well-designed adversarial attacks. A wide range of techniques have been proposed towards this area and achieved significant performance \cite{cai2005mining,goodfellow2014explaining,xu2019topology}. % \cite{DBLP:conf/kdd/ZugnerAG18} first propose an efficient attack performing perturbations on both graph structure and node features, and also making effort to ensure its ``unnoticeability''. In a follow-up study, 
For instance, \cite{DBLP:conf/iclr/ZugnerG19} study the discreteness of graph structure data and further propose a poisoning attack method applying meta learning. \cite{xu2019topology} presents a novel gradient-based method to inject edges from an optimization perspective, also displaying an impressive ability to fool the GNNs. 
So, in the consideration of all the redundant edges as well as the deliberately injected edges, we further raise the following questions: Can we find out these edges before the training process? Can GNNs keep their performance without all those deleted edges on clean graphs? Can we enhance the robustness of GNNs against adversarial attacks by deleting these edges?

%In fact, while most Graph Neural Networks (GNNs) operating on graph structure data tend to consider all of the edges in a graph to train its model \cite{wu2019simplifying,hamilton2017inductive,zhu2019robust,DBLP:conf/iclr/KipfW17}, there are more and more attempts trying to leave out edges during the training process, which may lead to deeper and more robust models against adversarial attacks \cite{DBLP:conf/iclr/RongHXH20,DBLP:journals/corr/abs-2010-11797}. The successful training and the enhanced performance of these models also happen to prove that not all edges are necessary for a task on graphs, especially graphs under well-designed attacks. In this spirit, we further raise the following questions: Can we find out these redundant or deliberately injected edges before the training process? Can GNNs keep their performance without all those deleted edges on clean graphs? Can we enhance the robustness of GNNs against adversarial attacks by deleting these edges?

In this work, we focus on the information and effects that edges bring to the graph and propose a \textbf{Unified Graph Purification (UGP)} framework to preprocess the data before the training of GNNs. Under the proposed UGP framework, the core problems are divided into evaluating each of the edges by intrinsic features of the data and deleting the redundant or misleading edges with the least loss of information. We use the powerful Graph Convolutional Network (GCN) proposed by \cite{DBLP:conf/iclr/KipfW17} as the estimating model, which, as a matter of fact, can be replaced by any other more accurate GNNs, and examine the performance of several measurements for edges implementing the UGP framework. We also introduce a residual-iteration (RI) strategy for measurements requiring unknown information like labels and thus needing a pre-trained model to predict such information. Experimental results show that the proposed framework applying the selected measurements can indeed delete edges and simultaneously keep the performance of GNNs. Further experiments also display the ability of these methods to enhance the robustness of GNNs against adversarial attacks.

Our main contributions exploring this area are summarized as follows:
\begin{itemize}
    \item We introduce a unified graph purification framework for detecting and purifying redundant and injected edges in a graph.
    \item We introduce a residual-iteration strategy to measurements requiring information unknown to the preprocess model and thus need a surrogate model to predict such information.
    \item We conduct experiments to show that the proposed framework can not only find out the redundant edges on clean graphs but also recognize the adversarial edges generated by malicious attackers and thus enhance the robustness of GNNs.
\end{itemize}

\section{Related Work}
With great attention dropped on the powerful GCNs, more and more researchers start to consider about leaving out certain edges of graph structure data and thus heading for more robust models. After \cite{DBLP:conf/iclr/VelickovicCCRLB18} first discuss applying dropout on edge attentions in GAT, \cite{DBLP:conf/iclr/RongHXH20} develop this idea into dropping edges randomly during training and present the formulation of DropEdge. When it comes to the inference stage, \cite{DBLP:journals/corr/abs-2010-11797} propose an adaptive inference mechanism, which drops out all the edges of a single node and only trusts its own features to formulate a counterfactual inference. The successful training of these models is telling us that, indeed, a huge number of edges are not necessary for the training and the inference stage of GNNs. However, these models dropout edges either in a completely random way or just delete all the edges, which, obviously, is missing a evaluation and selection process.

Moreover, a line of recent studies has demonstrated that GNN-based models suffer from vulnerabilities to adversarial perturbations due to strongly relying on the graph structure and local information \cite{chen2020survey}.  By injecting several poisoning data (i.e., adversarial edges) into the graph, attackers can easily fool the GNNs and cause significant degradation in node classification performance  \cite{10.1145/3219819.3220078,DBLP:conf/iclr/ZugnerG19}. The model trained on the perturbed graph suffers from the influence of noisy or adversarial data, thus restricting their application in real-world scenarios. Knowing the fact that GNNs are highly vulnerable against adversarial attacks especially with multiple edges injection, there is an urgent need to design practical methods to purify the graph and improve the robustness of GNNs.

In this vein, \cite{DBLP:conf/ijcai/Wu0TDLZ19} propose to discover adversarial edges via Jaccard similarity scores of the end nodes' features. \cite{zhang2019comparing} study the statistical differences between unperturbed graphs and perturbed graphs, and further propose to detect adversarial edges by calculating the Kullback-Leibler (KL) divergences \cite{Joyce2011} between the softmax probabilities of node and its neighbors. \cite{DBLP:conf/wsdm/EntezariADP20} explore the characteristic of the high-rank spectrum of perturbed graphs and vaccinate the GNNs with low-rank approximations. These pre-processing techniques have indeed enhanced the robustness of GNNs against adversarial attacks, however, these methods are insufficient in the face of stronger attacks since they can only explore a relatively small amount of adversarial edges, which means there are still several redundant edges being left out. More importantly, there lacks a unified framework to summarize all those methods and guide the development of new measurements.

\section{Notations and Preliminaries}
\subsection{Notations}
In this paper, we mainly study the task of semi-supervised node classification in an undirected, unweighted graph, and leave the discussion of other tasks for future exploration. We follow the widely used notation and represent a graph with N nodes as $G=(V,E)$, where $V=\{v_1,..., v_N\}$ is a finite set of vertices (nodes) and $E=\{e_1, ..., e_K\}$ is a finite set of links (edges). We use a matrix $X \in \mathrm{R}^{N \times D}$ to denote the $D$ dimensional node features and an adjacency matrix $A\in \{0,1\}^{N \times N}$ to represent the connections of node pairs in graph $G$, where $A_{u,v}=1$ means node $u$ and $v$ is connected while $A_{u,v} = 0$ otherwise. In addition, for the node classification task, we use $Y\in \{0,1\}^{N \times C}$ to denote a set of class labels where $C$ is the number of  classes and $Y_i$ denotes the ground-truth label of node $v_i$.

\subsection{Graph Purification}
Graph purification was first mentioned by \cite{DBLP:journals/corr/abs-2003-00653}, in their empirical study as defending inserted poisoning attacks by purifying the graph data before training the GNNs. We further develop this concept by emphasizing the effect of purification methods on clean graphs. In fact, under the real circumstances, the defenders are supposed to have absolutely no idea if a dataset is under attack or not, and it is not a common thing that the graph is perturbed. So, the top priority of purification methods should be keeping the models’ performance on clean data, which means the purification of edges should cost the least loss of information. And then, in this premise, find out as many redundant edges (on clean graphs) and poisons (on perturbed graphs) as possible.

\begin{figure*}[t]
\centering

\includegraphics[width=0.8\textwidth]{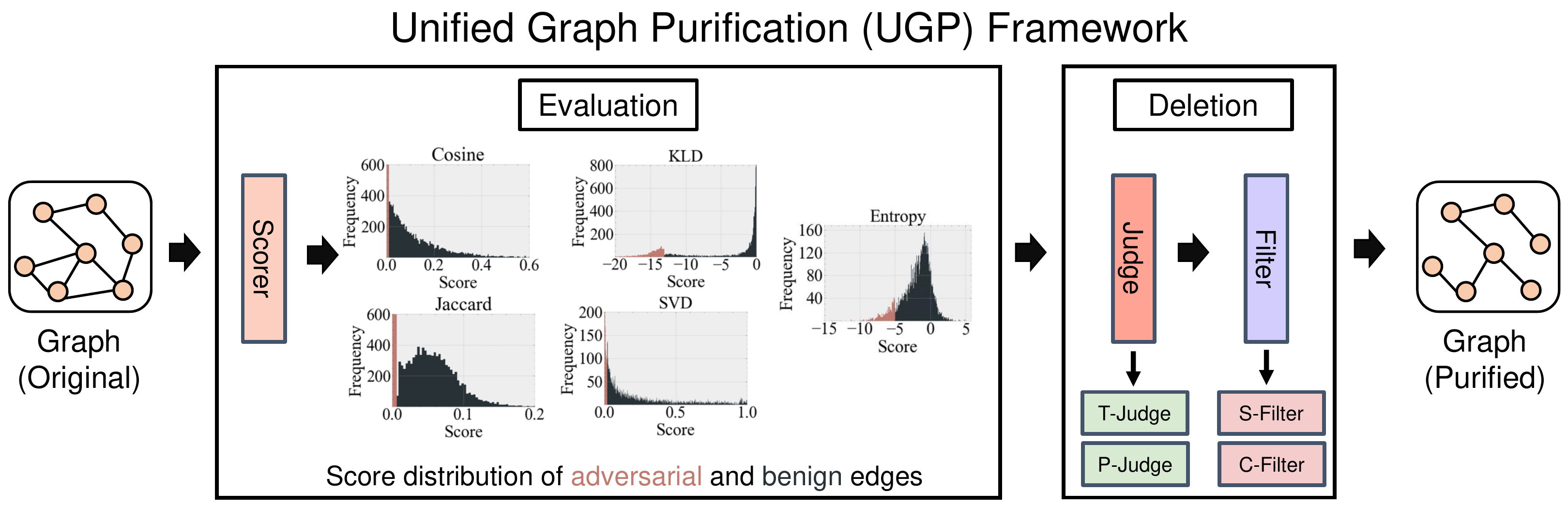}
\caption{Unified Graph Purification Framework. Scorer evaluates the edges and present the score distributions to Judge. Judge selects out the redundant edges according to the scores. Filter further checks these edges with certain schemes and leaves out the edges that may lead to loss of too much information without them.} \label{fig:ugp} 
% \vspace{-0.5cm}
\end{figure*}

\begin{figure}[t]
\centering
  \includegraphics[width=0.7\linewidth]{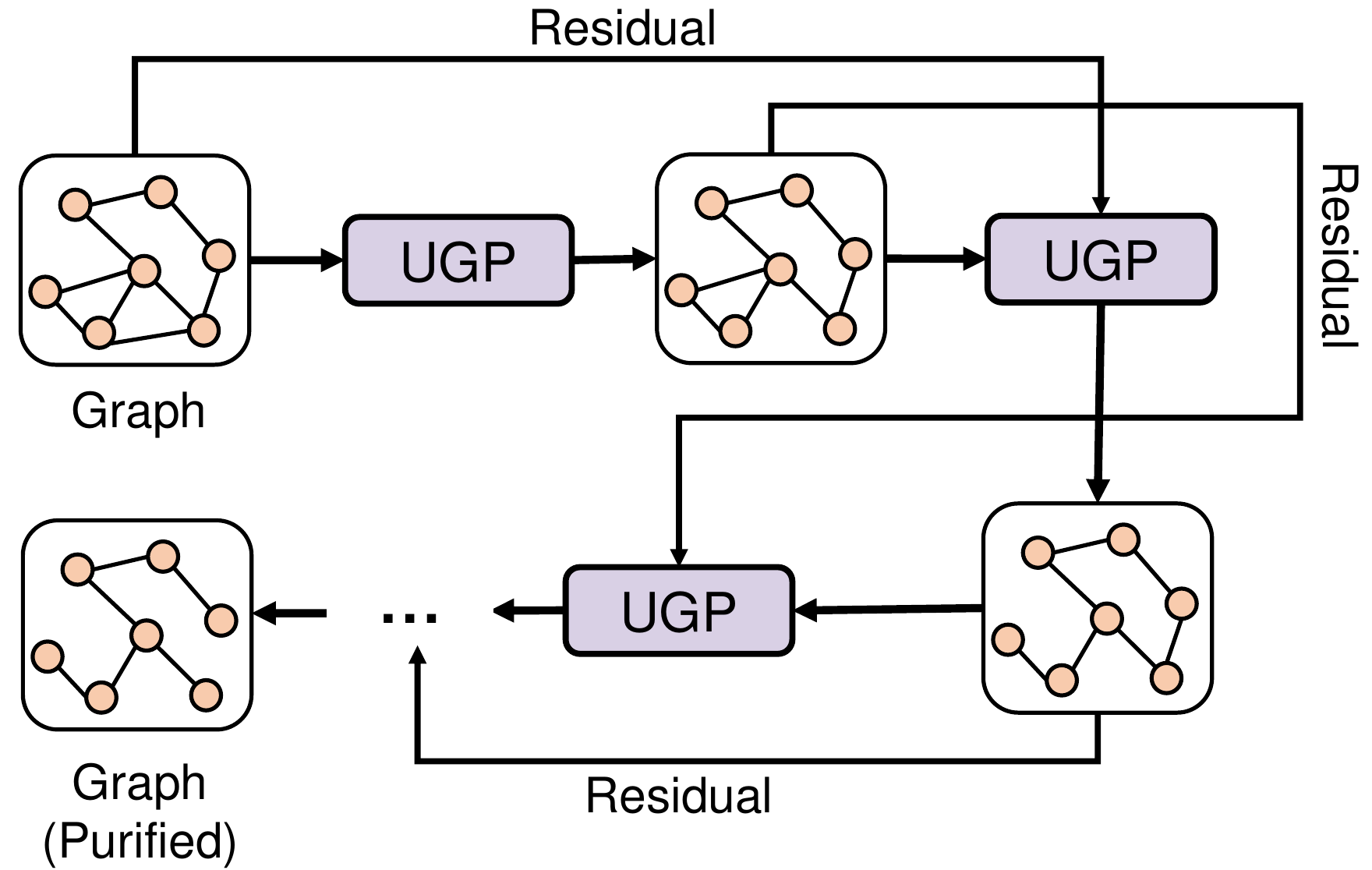}
\caption{The Residual-Iteration Strategy. The adjacent matrix of this iteration is sent to the next purification process as the residual and calculate scores to update the next adjacent matrix. In this way, the information lost through the last iteration can be somehow revisited in this iteration.} \label{fig:RI} 
% \vspace{-0.5cm}
\end{figure}

\section{Unified Graph Purification Framework}

In this section, we propose a unified framework of graph purification as depicted in Fig.~\ref{fig:ugp}. 

In Fig.~\ref{fig:ugp}, the Scorer is the measurement used to evaluate the edges. If the measurement only needs the features of nodes or the adjacent matrix of the graph (like Jaccard), then we can just calculate it directly and present it to Judge. However, when it comes to measurements requiring labels or other information that the defender has no access to, we then need an additional surrogate model to predict that information for us (like KL-divergence). In fact, the surrogate model can be set as any GNNs, as long as it provides the information that the Scorer needs. 

What’s more, since the GNNs are usually vulnerable to adversarial attacks, which means the prediction cannot be trusted at first, we introduce a residual-iteration strategy (depicted in Fig.~\ref{fig:RI}) for measurements requiring pre-trained surrogate models to solve this issue. Specifically, during the initial iterations, we delete the edges in a relatively conservative way. Then, while the model’s performance enhances little by little through the iterations, we gradually delete more and more edges utilizing more reliable information provided by the enhanced models. We stop the iteration when there are too few of edges left to delete or the performance on the validation set is not getting any better for a while, or just stop it after a given number of iterations. As for the residual part, we borrow the idea of ResNet \cite{DBLP:conf/cvpr/HeZRS16} and engage it in our framework. The residual block in ResNet is a shortcut connection for inputting the weights several layers before to the current layer and update together to solve the degradation issue. It has also been proved to have the power of refinement during the training process \cite{DBLP:journals/corr/LiaoP16,DBLP:journals/corr/GreffSS16,DBLP:journals/corr/abs-1710-04773}, which is exactly what we need, especially with the first few iterations. However, simply concatenating two matrices together doesn’t really fit in our framework. So, in the spirit of residual, we use the predicted information resulted from the current iteration and the adjacent matrix the earlier iteration instead to calculate scores and then update the current adjacent matrix. Ideally, this is supposed to perform a self-correction role through iterations, and the stronger the attacks are, the better the framework would perform with a residual operation. 

After Scorer gives out scores for the edges, we present the distribution of these scores to Judge, which will find out the redundant edges accordingly. Next, in order to delete edges with the least loss of information, Filter gives these edges one more check and leaves out the edges that may damage the original topological structure too much without them.

\subsection{Scorer}
To implement our framework, we summarize and further develop five preprocess measurements as Scorer, two of which need a pre-trained surrogate model through iterations. Most of these methods are based on two generally accepted empirical observations: a) Attackers usually tend to insert edges over removing them and b) Attackers are usually in favor of connecting edges between dissimilar nodes. 
Note that, %to make the scores more straight-forward and also consistent with each other, 
we construct the score of each edge by its extent of positive contribution to the whole graph given by the selected measurements. In other words, an edge is always more likely to be deleted with a lower score.

\begin{figure}
    %\centering
    \subfigure[Jaccard] {
        \label{fig3a}
        \includegraphics[width=0.45\linewidth]{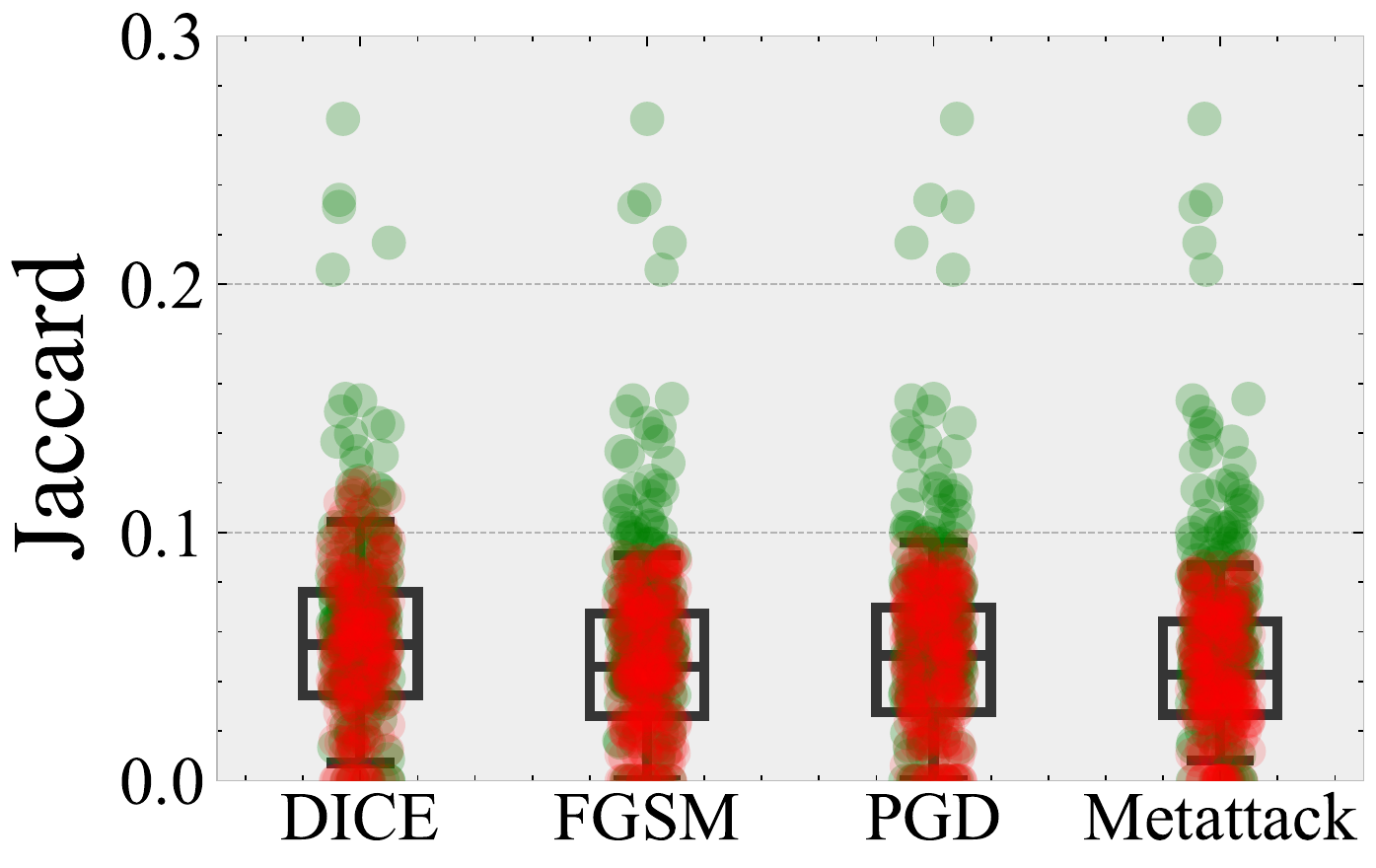}}
    \subfigure[Cosine] {
        \label{fig3b}
        \includegraphics[width=0.45\linewidth]{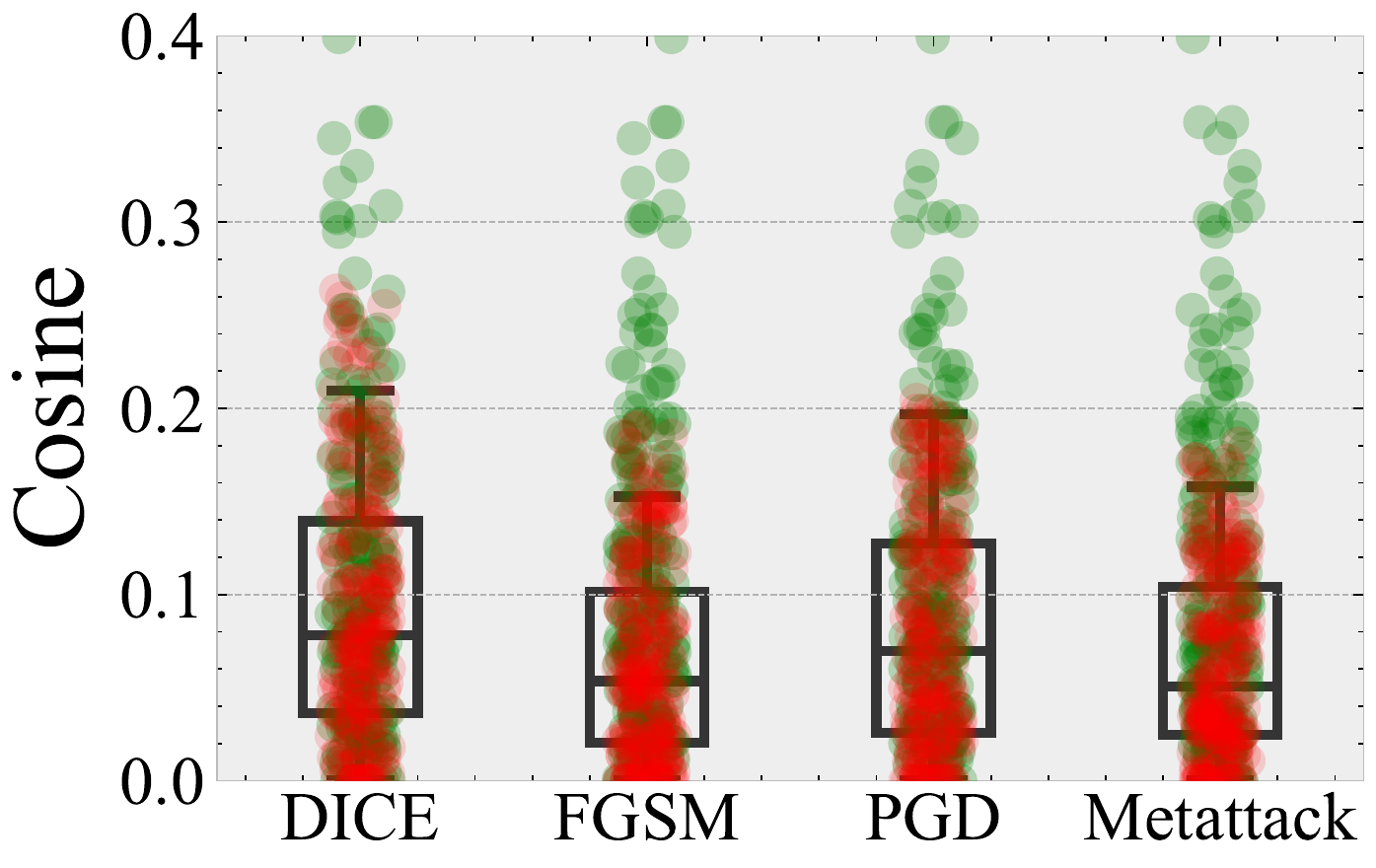}}\\
    \subfigure[SVD] {
        \label{fig3c}
        \includegraphics[width=0.45\linewidth]{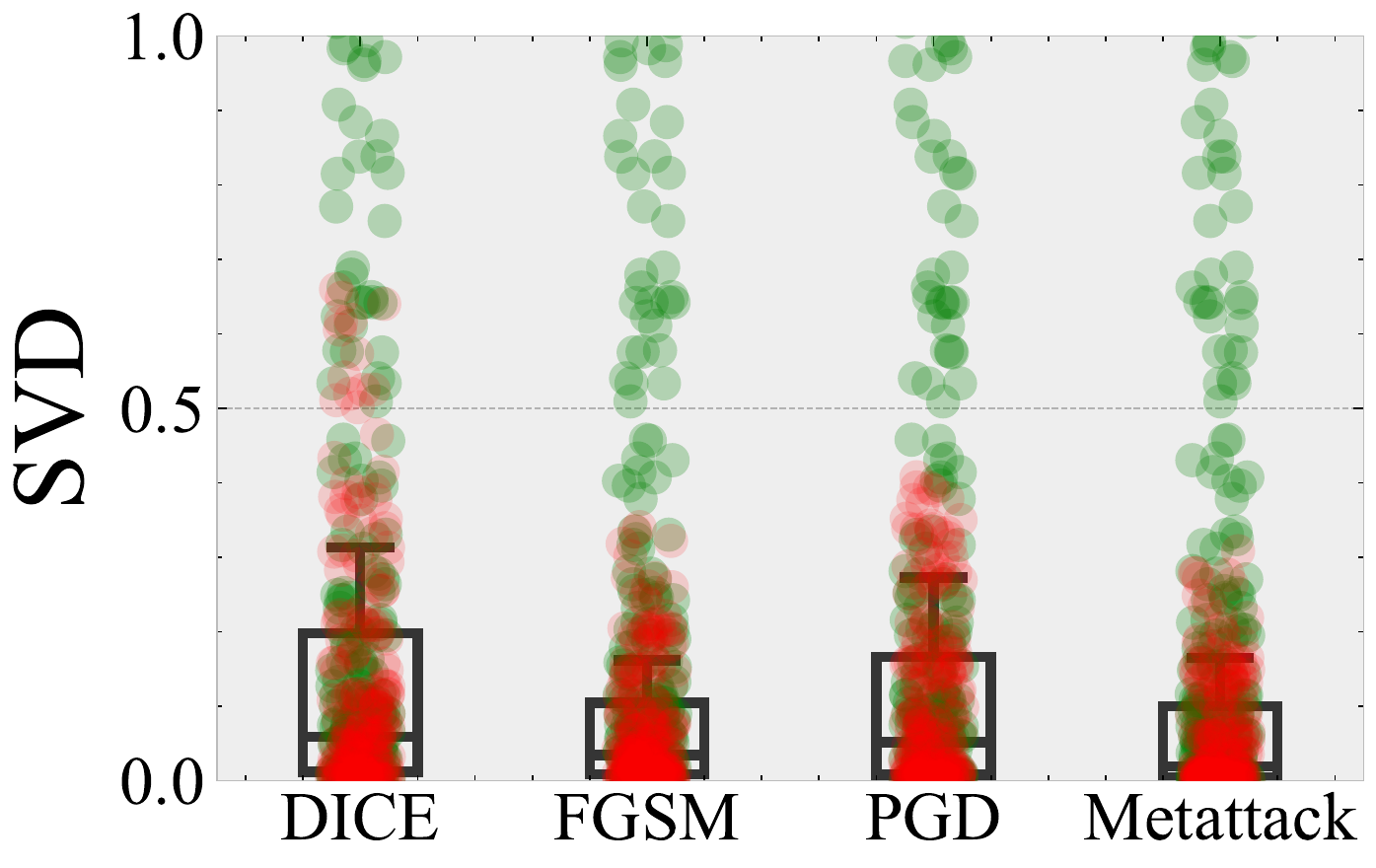}}
    \subfigure[KLD] {
        \label{fig3d}
        \includegraphics[width=0.45\linewidth]{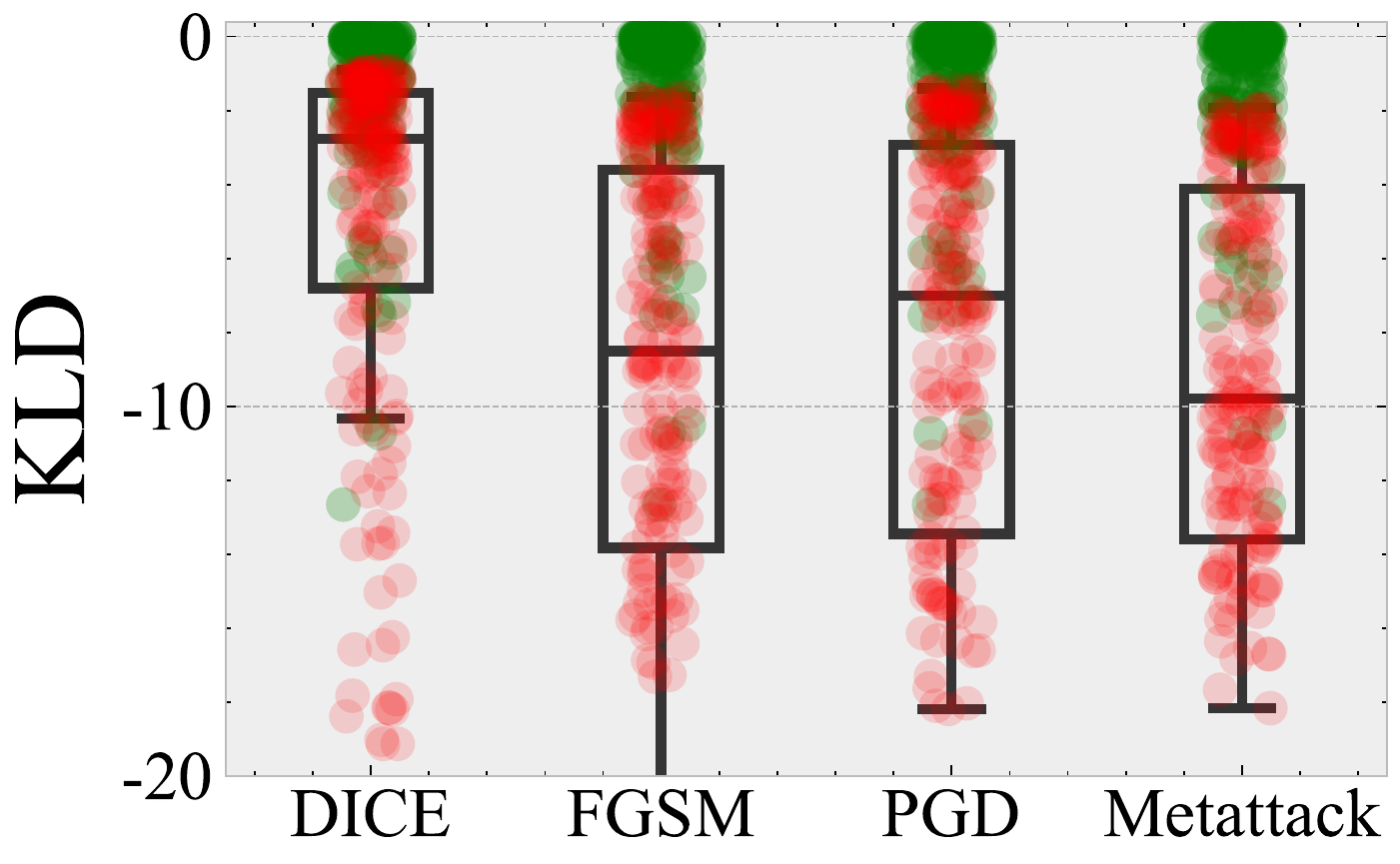}}\\
    \subfigure[Label Entropy] {
        \label{fig3e}
        \includegraphics[width=0.45\linewidth]{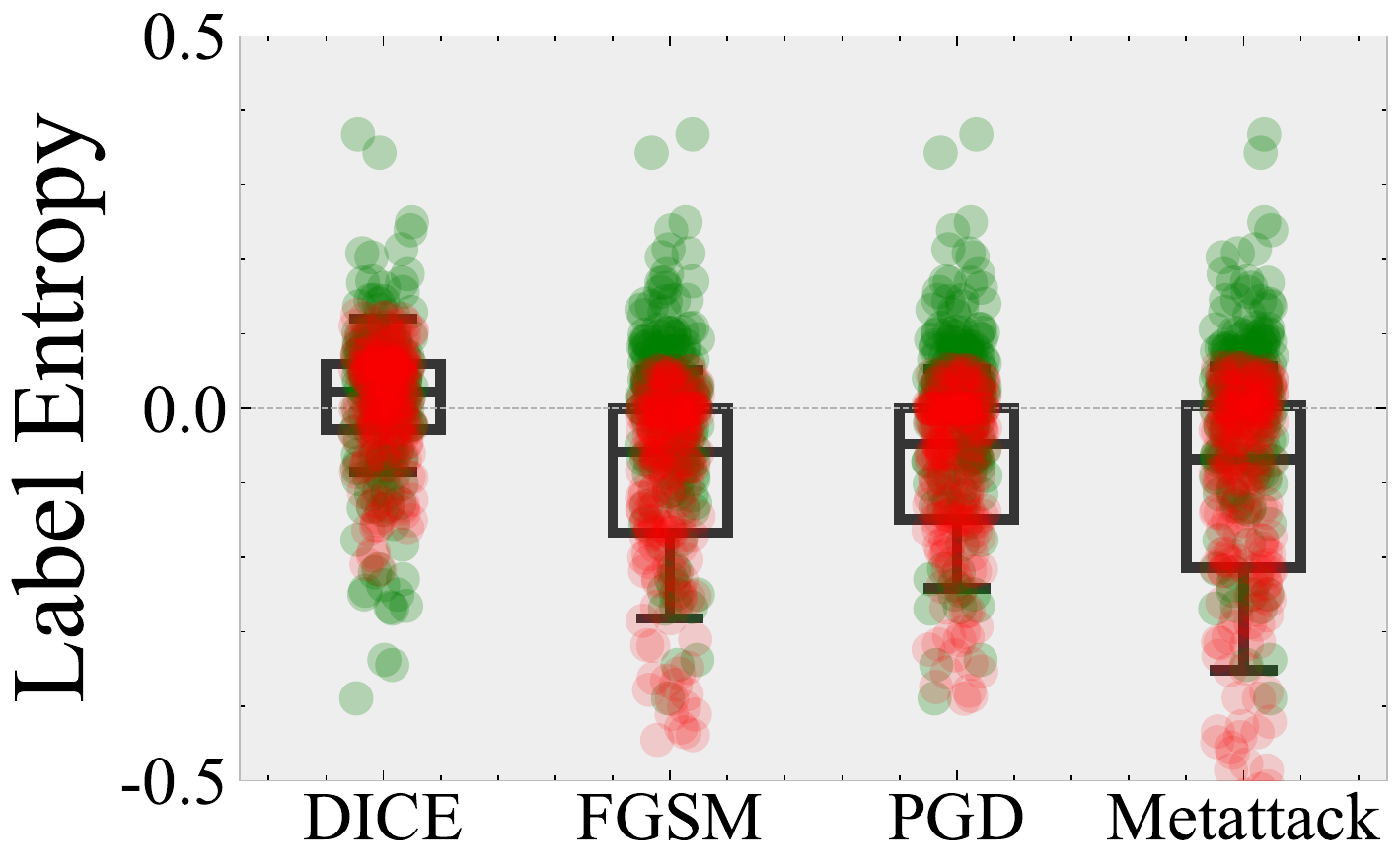}}
    \subfigure[Feature Entropy] {
        \label{fig3f}
        \includegraphics[width=0.45\linewidth]{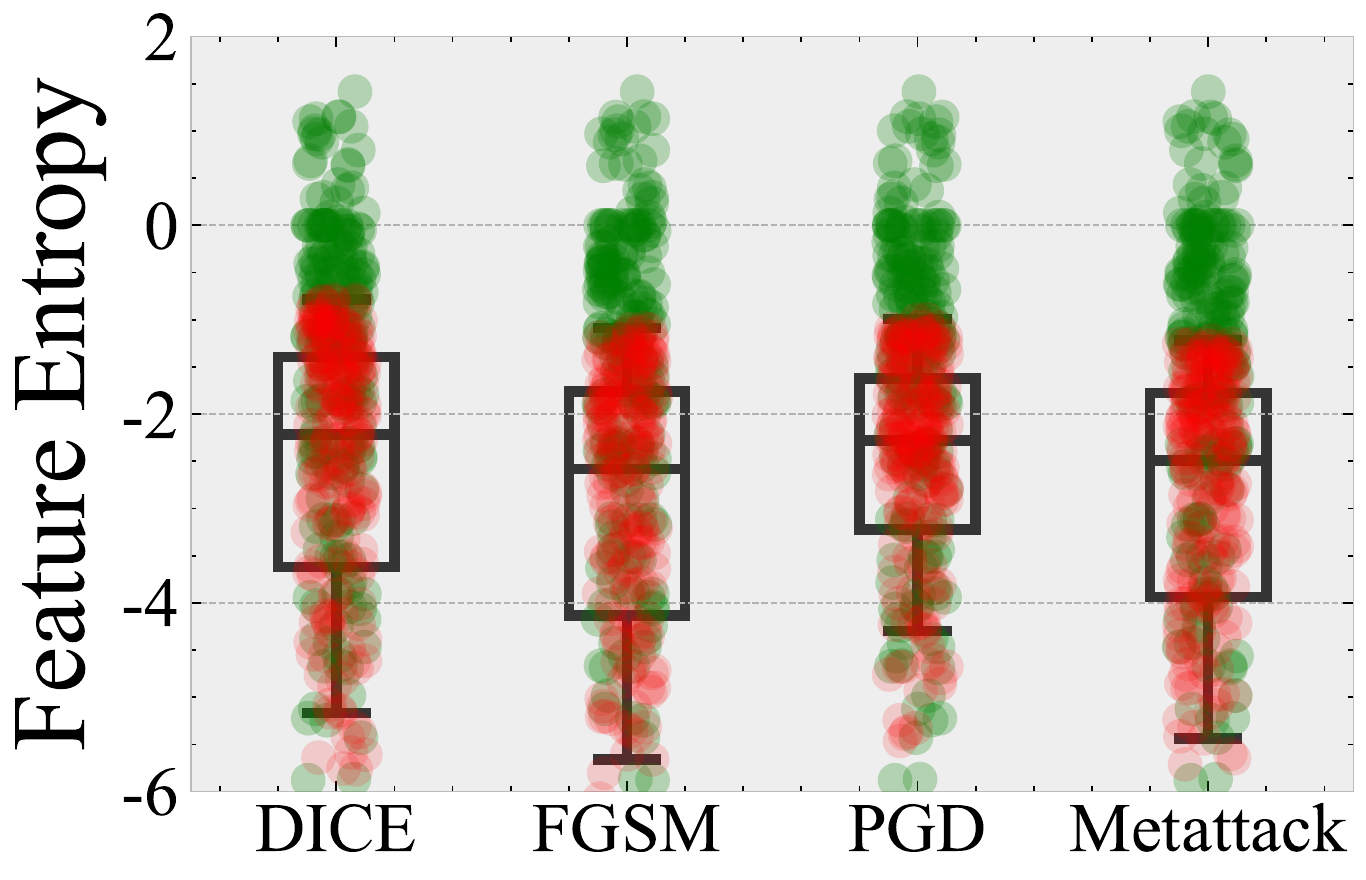}}
    \caption{Box plots of edges' score distributions for unperturbed graphs (green dots) and perturbed graphs (red dots, $\tau$=15\%) on the Cora-ml dataset.}
    \label{fig3}
\end{figure}

\subsubsection{Jaccard}
Jaccard Similarity has been adopted to discover adversarial edges \cite{DBLP:conf/ijcai/Wu0TDLZ19}. The key insight is that the adversarially manipulated graphs differ from normal graphs statistically especially in node similarity. Given a binary node feature matrix $X \in \{0,1\}^{N \times D}$, Jaccard similarity measures the overlap of features between node $u$ and the neighboring node $v$, the corresponding  score is calculated as follows
\begin{equation}
    \begin{aligned}
    S_{u,v}=&\frac{M_{1,1}}{M_{0,1}+M_{1,0}+M_{1,1}},\\
    M_{i,j}=&\sum_k^D (X_{u,k}=i\ \mbox{and}\ X_{v,k}=j), i,j \in \{0,1\}
    \end{aligned}
\end{equation}
Since Jaccard similarity score $S_{u, v}$ lies in the range of $[0,1]$, a larger value of this metric indicates the corresponding nodes are more similar.
\subsubsection{Cosine}
Since Jaccard similarity score is restricted for binary inputs, it is more reasonable to use Cosine similarity to wrestle with continuous data. Given a node feature matrix $X \in \mathrm{R}^{N \times D}$, Cosine similarity measures the feature similarity between node $u$ and the neighboring node $v$ by inner product space, and the corresponding score is calculated as follows
\begin{equation}
    \begin{aligned}
    S_{u,v}=&\frac{X_u \cdot X_v}{\|X_u\|\|X_v\|}
    \end{aligned}
\end{equation}
Note that a larger value of Cosine similarity score indicating two nodes are more similar.
\subsubsection{SVD}
\cite{DBLP:conf/wsdm/EntezariADP20} argues that the perturbed graph tends to be high-rank in the spectrum and leverage Singular Value Decomposition (SVD) and low-rank approximation of the graph to enhance its robustness. We extend this insight in our work by calculating the SVD score of each approximated edge
\begin{equation}
    \begin{aligned}
    S_{u,v}=\hat{A}_{u,v}, \mbox{where} \ \hat{A}=U \Sigma V^{T}
    \end{aligned}
\end{equation}
SVD score $S_{u, v}$ measures the connectivity of the edge $(u,v)$, which lies in the range of $[0,1]$, and a larger value of this metric indicates the corresponding edge (connection) has a higher probability to exist.

\subsubsection{Neighborhood Entropy} Entropy detection is based on the similar idea that attack methods tend to connect nodes with dissimilar features or different labels, and thus increase the entropy of features or labels of central nodes and their neighborhood. The calculation of entropy for a node is described as follow:
\begin{equation}
    \begin{aligned}
         & \mbox{NE}(u)     =-\sum_{l}^{L} P_{l}(u) \log \left(P_{d}(u)\right),                                                   \\
         & \mbox{where}\, P(u)  =\frac{p(u)}{\sum_{l}^{L} p_{l}(u)},                                                                      
    \end{aligned}
\end{equation}
where $L$ equals $D$ for feature entropy and equals $C$ for label entropy, and $p(u)$ for both entropy are calculated by:
\begin{equation}
    \begin{aligned}                                                                \\
         & p_{feature}(u)                 = \sum_{v \in \mathcal{N}(u) \cup u} \frac{X_{v}}{\sqrt{(|\mathcal{N}(u)|+1)}},     \\
         & p_{label}(u)                 = \sum_{v \in \mathcal{N}(u) \cup u} \frac{Y_{v}}{\sqrt{(|\mathcal{N}(u)|+1)}},
    \end{aligned}
\end{equation}
where $N(u)$ is the amount of neighbors for node $u$.
In fact, it is more of a nodewise measurement and we develop it into an evaluation for edges. Specifically, the score of an edge is the variation of entropy it brings to the nodes on both ends. In other words, the score for the edge $e$ between node $u$ and node $v$ is the difference between $NE(u) + NE(v)$ with and without edge $e$. We combine these two measurements together by first normalizing them and then add them together with the accuracy on the validation set or training set as a weight.

\subsubsection{KLD} KL-divergence (KLD) is originally a measurement based on the idea that attacks like Nettack \cite{10.1145/3219819.3220078} tends to create a discrepancy between the first-order proximity information of a node and that of its neighbors \cite{zhang2019comparing}. KLD evaluates each edge by calculating the KL divergence between the softmax probabilities of the nodes on both ends estimated by a surrogate model.
\begin{equation}
\begin{aligned}
&S_{u,v} = -K L(\hat{Y}_u \| \hat{Y}_v) - K L(\hat{Y}_v \| \hat{Y}_u),\\
&K L(\hat{Y}_u \| \hat{Y}_v)=\sum_k^C \hat{Y}_{u,k} \log (\frac{\hat{Y}_{u,k}}{\hat{Y}_{v,k}})
\end{aligned}
\end{equation}
where $\hat{Y}$ is the prediction of the surrogate model or the feature matrix.

Similar to what we did with label entropy, we further develop KLD by combining the feature with it as an adjustment, which would make this measurement not entirely dependent on the performance of the surrogate model and thus become more effective when the prediction is not that trustworthy. Specifically, we also calculated the KLD between $X_u$ and $X_v$ and add them together with the KLD between possibility vectors as the final score.

\subsection{Judge}
We propose two kinds of Judge to find out the redundant edges according to the scores.

\noindent \textbf{P-Judge} P stands for \textit{percentage}, which means it selects out a certain percentage of remaining edges. Supposedly, P-judge endows the whole process with better control of the amounts of edges getting deleted, and thus fits the measurements that fluctuate a lot and need multiple iterations.

\noindent \textbf{T-Judge} T stands for \textit{threshold}, which means it selects out edges with a score higher or lower than the given threshold. We believe T-judge should be combined with measurements that are relatively stable and more trustworthy, or it may delete too many useful edges.

\subsection{Filter}
We propose two kinds of Filter to make sure the methods purify graphs with the least loss of information. 

\noindent \textbf{S-Filter} S stands for \textit{singleton}, which means this Judge makes sure there are no single nodes that get completely disconnected from the whole graph. In fact, if a node gets singled out, there will be no neighborhood information to aggregate. Though it does cut out any perturbations and result in enhanced robustness against adversarial attacks, it also leaves out the useful information from neighbors and sacrifices the performance on clean graphs. 

\noindent \textbf{C-Filter} C stands for \textit{connectivity}, which means this Judge makes sure the purification methods don’t break the connection of the graph. In this way, C-Judge keeps the information of a graph under purification with a rather different perspective: as long as the nodes are still connected, no matter how many nodes are in between, there will still be chances to aggregate their features together and utilize it to train the GNNs or predict the labels. After all, just as \cite{granovetter1973} points out, ``weak ties'' can sometimes be the gamechanger. So we believe it is important to keep the connection of the graph for certain applications. 

Specifically, to maintain the connectivity with respect to the scores, C-Judge sets weights accordingly to each of the edges in the original unweighted graph, where all of the edges weight 1 equally. Then, the weight of the selected edges would be set to its score plus one, making the selected edges all weight more than one and are also sorted by their scores. Finally, we apply Prim's algorithm \cite{6773228} on the weighted graph and find the minimum spanning tree (MST), which is the least cost to maintain the connectivity considering the scores. Leaving out the edges in the MST, C-Judge goes ahead and deletes the rest of the selected edges.

\section{Experiments and Discussions}

\subsection{Datasets and Setup}
In this paper, we adopt three commonly used datasets as benchmarks: Citeseer, Cora and Cora-ml \cite{sen2008collective}. For each dataset, we randomly select 20\% of the nodes to constitute the training set (10\% of which is set to be the validation set) and treat the rest of the nodes as the test set. Table~\ref{dataset}. shows an overview of the datasets.

As for the attack methods, we adopt a number of structure attack methods including state-of-the-arts to study the defense performance of these purification measurements: DICE \cite{cai2005mining}, FGSM \cite{goodfellow2014explaining}, PGD \cite{xu2019topology}, Metattack \cite{DBLP:conf/iclr/ZugnerG19}.

\begin{table}[t]
    \centering
    \begin{tabular}{c|c|c|c}
    \hline
        Dataset & \#Nodes & \#Edges & Density \\ \hline
        Citeseer & 2,110 & 3,668 & 0.082\% \\
        Cora & 2,485 & 5,069 & 0.082\% \\ 
        Cora-ml & 2,810 & 7,981 & 0.101\% \\ \hline        
    \end{tabular}
\caption{Dataset statistics. Joining previous works' practice, we only consider the largest connected component of the graph for each dataset.}	
\label{dataset}    
\end{table}

\subsection{Results and Discussions}
Given the experimental setup presented in the earlier section, we utilize the results of designed experiments to answer the following research questions:
\paragraph{RQ1} Can the implemented methods under the UGP framework generally enhance the performance of GNNs?
\paragraph{RQ2} How does each of the module in the UGP framework help?

\begin{table*}[]
    \resizebox{\textwidth}{!}{
        \begin{tabular}{l|lllllllll}
            \toprule
        \textbf{Dataset}  & \textbf{Non} & \textbf{Cosine} & \textbf{Jaccard} & \textbf{SVD} & \textbf{Entropy} & \textbf{I-Entropy} & \textbf{RI-Entropy} & \textbf{KLD} & \textbf{I-KLD} \\ 
            \midrule
        \textbf{Citeseer} & 71.15$\pm$0.17   & 71.09$\pm$0.13      & 71.09$\pm$0.13       & 70.02$\pm$0.37  & 70.97$\pm$0.31       & 71.02$\pm$0.34         & 70.68$\pm$0.25          & 71.03$\pm$0.34   & 69.91$\pm$0.21     \\
        \textbf{Cora}     & 83.35$\pm$0.39   & 83.30$\pm$0.34       & 83.30$\pm$0.34        & 78.47$\pm$0.42   & 83.45$\pm$0.26       & 83.70$\pm$0.42          & 83.65$\pm$0.13          & 82.9$\pm$0.09    & 81.94$\pm$0.30     \\
        \textbf{Cora-ml}  & 86.39$\pm$0.17   & 85.50$\pm$0.26       & 85.54$\pm$0.22       & 83.72$\pm$0.22   & 85.77$\pm$0.19       & 85.59$\pm$0.14         & 85.54$\pm$0.25          & 85.54$\pm$0.25   & 85.59$\pm$0.23     \\
            \bottomrule
        \end{tabular}
    }
    \caption{Accuracy (\%) of different purification methods on clean graphs. RI-Entropy means implementing Entropy with residual-iteration strategy, while I-KLD and I-Entropy mean the implementations do iterations without residual operation. }   
	\label{tab2}
\end{table*}

\begin{table*}[]
    \resizebox{\textwidth}{!}{
\begin{tabular}{l|l|l|l|l|l|l|l|l|l}
     \toprule
     & \textbf{Non}      & \textbf{Jaccard-TS} & \textbf{Jaccard-PS} & \textbf{Jaccard-TC} & \textbf{Jaccard-PC} & \textbf{KLD-TS}     & \textbf{KLD-PS}     & \textbf{KLD-TC}     & \textbf{KLD-PC}     \\
     \midrule
\textbf{Clean}   & 86.17$\pm$0.17 & 85.43$\pm$0.09 & 84.85$\pm$0.29 & 86.12$\pm$0.17 & 85.92$\pm$0.11 & 86.09$\pm$0.21 & 84.79$\pm$0.36 & \textbf{86.25$\pm$0.18} & 85.3$\pm$0.35  \\
     \midrule
\textbf{1\%} & 85.44$\pm$0.25 & 84.92$\pm$0.18 & 84.46$\pm$0.23 & 85.25$\pm$0.3  & 85.29$\pm$0.15 & 85.7$\pm$0.12  & 84.55$\pm$0.21 & \textbf{86.09$\pm$0.21} & 85.33$\pm$0.26 \\
\textbf{5\%} & 79.56$\pm$0.23 & 79.84$\pm$0.4  & 82.0$\pm$0.11  & 79.55$\pm$0.36 & 79.71$\pm$0.37 & \textbf{84.65$\pm$0.2}  & 84.38$\pm$0.17 & 84.14$\pm$0.11 & 83.92$\pm$0.28 \\
\textbf{10\%}  & 74.0$\pm$0.28  & 75.55$\pm$0.42 & 78.69$\pm$0.31 & 74.13$\pm$0.28 & 74.12$\pm$0.34 & 83.76$\pm$0.18 & \textbf{83.99$\pm$0.26} & 83.15$\pm$0.13 & 82.96$\pm$0.16 \\
\textbf{15\%} & 69.6$\pm$0.53  & 71.19$\pm$0.48 & 75.84$\pm$0.33 & 70.24$\pm$0.57 & 70.43$\pm$0.56 & 83.17$\pm$0.25 & \textbf{83.6$\pm$0.16}  & 81.81$\pm$0.23 & 82.22$\pm$0.2  \\
\textbf{20\%}  & 64.2$\pm$0.91  & 66.12$\pm$0.42 & 71.74$\pm$0.45 & 65.07$\pm$0.95 & 65.18$\pm$0.77 & \textbf{81.09$\pm$0.15} & 79.69$\pm$0.2  & 80.12$\pm$0.31 & 78.78$\pm$0.09 \\
\textbf{25\%} & 56.21$\pm$0.4  & 58.72$\pm$0.44 & 67.19$\pm$0.54 & 57.83$\pm$0.48 & 57.79$\pm$0.85 & \textbf{73.88$\pm$0.47} & 69.95$\pm$0.43 & 73.42$\pm$0.42 & 69.64$\pm$0.57\\
\bottomrule
\end{tabular}
}
    \caption{Accuracy (\%) of different methods on perturbed graphs generated by Metattack. ``T'' and ``P'' stand for T-Judge and P-Judge, while ``S'' and ``C'' stand for S-Filter and C-Filter. For instance, Jaccard-TC stands for the methods applying Jaccard as the measurement for Scorer, selecting out edges with higher scores than a certain \textit{threshold}, and leaving out the edges that may damage to the \textit{connectivity} of the graph.}   
	\label{tab3}
\end{table*}
\subsection{RQ1: Enhanced Performance}

To study and compare the performance between different proposed measurements, we train a GCN in the end to evaluate the final purification performance of the measurements and display the accuracy on purified clean graph in Table~\ref{tab2}. We also adopt several attack methods to poison the input graph and evaluate the methods' performance against adversarial attacks. We only report the plots of perturbed graphs attacked by Metattack and FGSM with different perturbation rates on Fig~\ref{fig3}, and leave other figures in the supplementary material. 
%In addition, we also introduce a measurement for robustness calculated by $\sum_i \tau_i p_i$, where $\tau_i$ is different perturbation rate of attacks and $p_i$ is the corresponding accuracy of GCN on the purified graph. We display the robustness performance of each implementation under different attacks on Table~\ref{tab2}. Note that the iteration times are set to be six in our experiments. 

\begin{figure}
    %\centering
    %\subfigure[Citeseer] {
    %    \label{fig3a}
    %    \includegraphics[width=0.5\linewidth]{citeseer_Metattack.png}}
    \subfigure[Cora(Metattack)] {
        \label{fig4a}
        \includegraphics[width=0.45\linewidth]{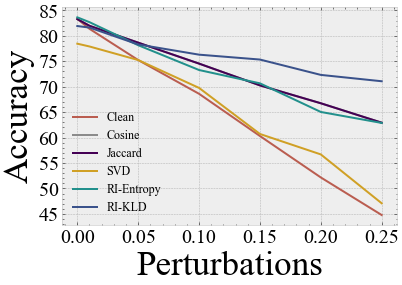}}
    \subfigure[Cora-ml(Metattack)] {
        \label{fig4b}
        \includegraphics[width=0.45\linewidth]{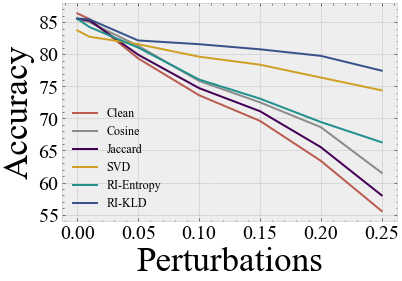}}\\
    
    \subfigure[Cora(FGSM)] {
        \label{fig4c}
        \includegraphics[width=0.45\linewidth]{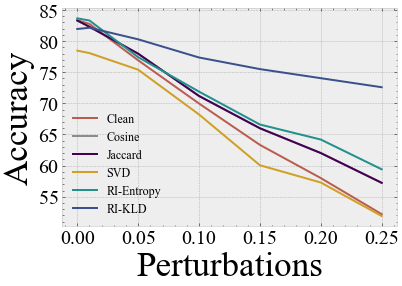}}
    \subfigure[Cora-ml(FGSM)] {
        \label{fig4d}
        \includegraphics[width=0.45\linewidth]{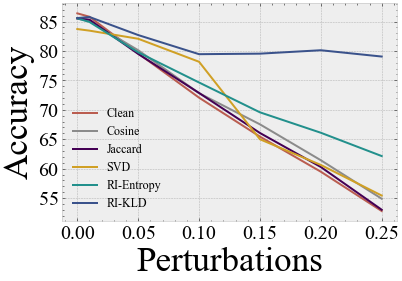}}
    
    \caption{Plots of performance under Metattack and FGSM with different perturbations rates on Cora-ml and Cora.}
    \label{fig4}
\end{figure}

% Observed from Table~\ref{tab2}, all of the purification measurements cannot surpass the performance on clean graph without purification. However, we follow previous works’ practice and consider any decrease under 2\% not significant or, in other words, an acceptable loss of information. In this case, only

It is observed that almost every measurement implementing UGP framework can enhance the robustness of the final GCN model, and only SVD betrays a significantly poorer performance on clean graphs (decrease over 2\%) and thus cause the loss of too much information. In this case, we don't consider SVD a qualified graph purification method, despite its relatively high robustness against adversarial attacks. On the other hand, the best robustness performance is mostly achieved by our proposed method I-KLD, especially for strong attacks like Metattack with high perturbation rates. This is because the iteration strategy engaged in I-KLD can help it find out more edges through iterations comparing with methods like Jaccard and Consine. Moreover, KLD is capable of utilizing the whole proximity information of each node instead of the hard prediction with the highest possibility, which further gives this measurement a higher tolerance to the poor performance of the surrogate model comparing with Entropy. 

\subsection{RQ2: Modules in the UGP Framework}

\begin{figure}
    %\centering
    %\subfigure[Citeseer] {
    %    \label{fig4a}
    %    \includegraphics[width=0.5\linewidth]{citeseer_Metattack_0.25.png}}
    \subfigure[Cora] {
        \label{fig5a}
        \includegraphics[width=0.45\linewidth]{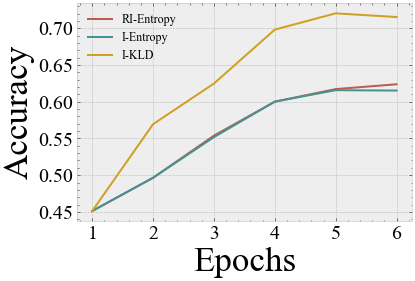}}
    \subfigure[Citeseer] {
        \label{fig5b}
        \includegraphics[width=0.45\linewidth]{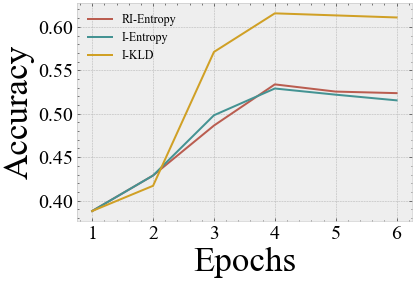}}
    \caption{Trends of accuracy on the test set enhancing through iterations provided by measurements with iteration strategy. The attack here is generated by Metattack with a perturbation rate of 0.25.}
    \label{fig5}
\end{figure}

In order to show the effectiveness of the selected measurements, we present the score distributions for edges calculated by each measurement on Cora-ml by Fig~\ref{fig3}. As for the effect of our proposed residual-iteration strategy, we display the trends of accuracy on test set through iterations provided by RI-Entropy, I-Entropy and I-KLD in Fig.~\ref{fig5}. We also construct experiments to demonstrate the effect of each of our proposed Judge and Filter by engaging Jaccard and KLD with different combinations of each modules and display the results in Table.~\ref{tab3}.

In Fig~\ref{fig3}, a clear difference between the scores for the unperturbed versus perturbed nodes ($\tau$=15\%) can be observed, especially for relatively stronger attacks like Metattack. So the deletion of such edges with relatively lower scores is supposed to enhance the performance of GNNs against adversarial attacks, which is exactly what happened in Fig~\ref{fig4}. And, naturally, the more green nodes got separated from the red nodes, the better the measurement is. The residual-iteration strategy is also working as expected, as the accuracy is generally increasing with every iteration and the methods applying the residual strategy outperforms the one without it. 
%Especially when attacks are strong, which means the accuracy of the initial surrogate model is quite low, the residual strategy indeed performs a self-corrected effect on the UGP framework. 
What's more, observed from Table~\ref{tab3}, Jaccard-TC and KLD-TC achieve the best performance comparing with methods utilizing the same measurement, and Jaccard-PS achieve the highest robustness compared to other Jaccard methods. This is because T-Judge and C-Filter play a similar role by conserving information on clean graphs and perturbed graphs with low perturbation rates, while P-Judge and S-Filter endow the methods with the ability to delete more edges and thus resulted in high robustness against strong attacks. As for KLD methods engaging T-Judge outperforming the ones with P-Judge, it is because we set the \textit{threshold} empirically according to this exact dataset. However, the distribution of scores calculated by KLD would not be the same on different datasets, and in real-life circumstances, we normally don't have the access to such information. In this case, we don't recommend T-Judge for measurements like KLD.

\section{Conclusion and Future Work}

In this work, we further clarify the concept of graph purification and present the UGP framework, a novel framework to preprocess the graph structure data before the training of GNNs starts. We implement the UGP framework by several measurements and construct experiments on three datasets with several attack algorithms to show that we can indeed enhance the robustness of GNNs against adversarial attacks through graph purification while not sacrificing the performance on clean graphs, and our proposed I-KLD performs the best in general consideration. It is expected that our research can offer a new perspective towards preprocessing graph structure data and more robust GNNs.

One potential direction for future work is to use machine-learning models instead of pure statistic measurements to evaluate each of the edges. Moreover, GNNs can keep their performance without that many edges may also indicate that current models couldn’t make use of all of the information provided in a graph, and thus encourages us to look for other means to utilize these deleted edges and improve the overall performance of GNNs.

\bibliographystyle{named}
\bibliography{ijcai21}

\begin{thebibliography}{}

\bibitem[\protect\citeauthoryear{Boshmaf \bgroup \em et al.\egroup
  }{2011}]{10.1145/2076732.2076746}
Yazan Boshmaf, Ildar Muslukhov, Konstantin Beznosov, and Matei Ripeanu.
\newblock The socialbot network: When bots socialize for fame and money.
\newblock ACSAC '11, page 93–102. Association for Computing Machinery, 2011.

\bibitem[\protect\citeauthoryear{Cai \bgroup \em et al.\egroup
  }{2005}]{cai2005mining}
Deng Cai, Zheng Shao, Xiaofei He, Xifeng Yan, and Jiawei Han.
\newblock Mining hidden community in heterogeneous social networks.
\newblock In {\em Proceedings of the 3rd international workshop on Link
  discovery}, pages 58--65. ACM, 2005.

\bibitem[\protect\citeauthoryear{Chen \bgroup \em et al.\egroup
  }{2018}]{chen2018heterogeneous}
Liang Chen, Yang Liu, Zibin Zheng, and Philip Yu.
\newblock Heterogeneous neural attentive factorization machine for rating
  prediction.
\newblock In {\em CIKM}, pages 833--842. ACM, 2018.

\bibitem[\protect\citeauthoryear{Chen \bgroup \em et al.\egroup
  }{2019}]{DBLP:conf/ijcai/ChenL0GZ19}
Liang Chen, Yang Liu, Xiangnan He, Lianli Gao, and Zibin Zheng.
\newblock Matching user with item set: Collaborative bundle recommendation with
  deep attention network.
\newblock In {\em IJCAI}, pages 2095--2101, 2019.

\bibitem[\protect\citeauthoryear{Chen \bgroup \em et al.\egroup
  }{2020}]{chen2020survey}
Liang Chen, Jintang Li, Jiaying Peng, Tao Xie, Zengxu Cao, Kun Xu, Xiangnan He,
  and Zibin Zheng.
\newblock A survey of adversarial learning on graph.
\newblock {\em arXiv preprint arXiv:2003.05730}, 2020.

\bibitem[\protect\citeauthoryear{Entezari \bgroup \em et al.\egroup
  }{2020}]{DBLP:conf/wsdm/EntezariADP20}
Negin Entezari, Saba~A. Al{-}Sayouri, Amirali Darvishzadeh, and Evangelos~E.
  Papalexakis.
\newblock All you need is low (rank): Defending against adversarial attacks on
  graphs.
\newblock In {\em WSDM}, pages 169--177, 2020.

\bibitem[\protect\citeauthoryear{Feng \bgroup \em et al.\egroup
  }{2020}]{DBLP:journals/corr/abs-2010-11797}
Fuli Feng, Weiran Huang, Xin Xin, Xiangnan He, and Tat{-}Seng Chua.
\newblock Should graph convolution trust neighbors? {A} simple causal inference
  method.
\newblock {\em CoRR}, abs/2010.11797, 2020.

\bibitem[\protect\citeauthoryear{Goodfellow \bgroup \em et al.\egroup
  }{2015}]{goodfellow2014explaining}
Ian~J. Goodfellow, Jonathon Shlens, and Christian Szegedy.
\newblock Explaining and harnessing adversarial examples.
\newblock In {\em ICLR}, 2015.

\bibitem[\protect\citeauthoryear{Granovetter}{1973}]{granovetter1973}
M.S. Granovetter.
\newblock {The Strength of Weak Ties}.
\newblock {\em The American Journal of Sociology}, 78(6):1360--1380, 1973.

\bibitem[\protect\citeauthoryear{Greff \bgroup \em et al.\egroup
  }{2016}]{DBLP:journals/corr/GreffSS16}
Klaus Greff, Rupesh~Kumar Srivastava, and J{\"{u}}rgen Schmidhuber.
\newblock Highway and residual networks learn unrolled iterative estimation.
\newblock {\em CoRR}, abs/1612.07771, 2016.

\bibitem[\protect\citeauthoryear{He \bgroup \em et al.\egroup
  }{2016}]{DBLP:conf/cvpr/HeZRS16}
Kaiming He, Xiangyu Zhang, Shaoqing Ren, and Jian Sun.
\newblock Deep residual learning for image recognition.
\newblock In {\em {CVPR} 2016}, pages 770--778. {IEEE} Computer Society, 2016.

\bibitem[\protect\citeauthoryear{Jastrzebski \bgroup \em et al.\egroup
  }{2017}]{DBLP:journals/corr/abs-1710-04773}
Stanislaw Jastrzebski, Devansh Arpit, Nicolas Ballas, Vikas Verma, Tong Che,
  and Yoshua Bengio.
\newblock Residual connections encourage iterative inference.
\newblock {\em CoRR}, abs/1710.04773, 2017.

\bibitem[\protect\citeauthoryear{Jin \bgroup \em et al.\egroup
  }{2020}]{DBLP:journals/corr/abs-2003-00653}
Wei Jin, Yaxin Li, Han Xu, Yiqi Wang, and Jiliang Tang.
\newblock Adversarial attacks and defenses on graphs: {A} review and empirical
  study.
\newblock {\em CoRR}, abs/2003.00653, 2020.

\bibitem[\protect\citeauthoryear{Joyce}{2011}]{Joyce2011}
James~M. Joyce.
\newblock {\em Kullback-Leibler Divergence}, pages 720--722.
\newblock Springer Berlin Heidelberg, Berlin, Heidelberg, 2011.

\bibitem[\protect\citeauthoryear{Kipf and
  Welling}{2017}]{DBLP:conf/iclr/KipfW17}
Thomas~N. Kipf and Max Welling.
\newblock Semi-supervised classification with graph convolutional networks.
\newblock In {\em ICLR}, 2017.

\bibitem[\protect\citeauthoryear{Liao and
  Poggio}{2016}]{DBLP:journals/corr/LiaoP16}
Qianli Liao and Tomaso~A. Poggio.
\newblock Bridging the gaps between residual learning, recurrent neural
  networks and visual cortex.
\newblock {\em CoRR}, abs/1604.03640, 2016.

\bibitem[\protect\citeauthoryear{{Prim}}{1957}]{6773228}
R.~C. {Prim}.
\newblock Shortest connection networks and some generalizations.
\newblock {\em The Bell System Technical Journal}, 36(6):1389--1401, 1957.

\bibitem[\protect\citeauthoryear{Rong \bgroup \em et al.\egroup
  }{2020}]{DBLP:conf/iclr/RongHXH20}
Yu~Rong, Wenbing Huang, Tingyang Xu, and Junzhou Huang.
\newblock Dropedge: Towards deep graph convolutional networks on node
  classification.
\newblock In {\em {ICLR}0}. OpenReview.net, 2020.

\bibitem[\protect\citeauthoryear{Sen \bgroup \em et al.\egroup
  }{2008}]{sen2008collective}
Prithviraj Sen, Galileo Namata, Mustafa Bilgic, Lise Getoor, Brian Galligher,
  and Tina Eliassi-Rad.
\newblock Collective classification in network data.
\newblock {\em AI magazine}, 29(3):93--93, 2008.

\bibitem[\protect\citeauthoryear{Velickovic \bgroup \em et al.\egroup
  }{2018}]{DBLP:conf/iclr/VelickovicCCRLB18}
Petar Velickovic, Guillem Cucurull, Arantxa Casanova, Adriana Romero, Pietro
  Li{\`{o}}, and Yoshua Bengio.
\newblock Graph attention networks.
\newblock In {\em ICLR}, 2018.

\bibitem[\protect\citeauthoryear{Wu \bgroup \em et al.\egroup
  }{2019}]{DBLP:conf/ijcai/Wu0TDLZ19}
Huijun Wu, Chen Wang, Yuriy Tyshetskiy, Andrew Docherty, Kai Lu, and Liming
  Zhu.
\newblock Adversarial examples for graph data: Deep insights into attack and
  defense.
\newblock In {\em IJCAI}, pages 4816--4823, 2019.

\bibitem[\protect\citeauthoryear{Xu \bgroup \em et al.\egroup
  }{2019}]{xu2019topology}
Kaidi Xu, Hongge Chen, Sijia Liu, Pin{-}Yu Chen, Tsui{-}Wei Weng, Mingyi Hong,
  and Xue Lin.
\newblock Topology attack and defense for graph neural networks: An
  optimization perspective.
\newblock In {\em IJCAI}, pages 3961--3967, 2019.

\bibitem[\protect\citeauthoryear{Zhang \bgroup \em et al.\egroup
  }{2019}]{zhang2019comparing}
Yingxue Zhang, S~Khan, and Mark Coates.
\newblock Comparing and detecting adversarial attacks for graph deep learning.
\newblock 2019.

\bibitem[\protect\citeauthoryear{Z{\"{u}}gner and
  G{\"{u}}nnemann}{2019}]{DBLP:conf/iclr/ZugnerG19}
Daniel Z{\"{u}}gner and Stephan G{\"{u}}nnemann.
\newblock Adversarial attacks on graph neural networks via meta learning.
\newblock In {\em ICLR}, 2019.

\bibitem[\protect\citeauthoryear{Z\"{u}gner \bgroup \em et al.\egroup
  }{2018}]{10.1145/3219819.3220078}
Daniel Z\"{u}gner, Amir Akbarnejad, and Stephan G\"{u}nnemann.
\newblock Adversarial attacks on neural networks for graph data.
\newblock KDD '18, page 2847–2856, 2018.

\end{thebibliography}
\end{document}